\documentclass{iopart}
\usepackage[dvips]{graphicx}
\begin{document}
\title[A stochastic CA for traffic flow with multiple metastable states]{A 
stochastic cellular automaton model for traffic flow with multiple 
metastable states}

\author{Katsuhiro Nishinari$^{1}${
\footnote{E-mail:knishi@rins.ryukoku.ac.jp}},
Minoru Fukui$^{2}${
\footnote{E-mail:fukui3@cc.nagoya-u.ac.jp}}
and Andreas Schadschneider$^{3}${
\footnote{E-mail:as@thp.uni-koeln.de}}
}
\address{$^1$\ Department of Applied Mathematics and Informatics, 
Ryukoku University, Shiga 520-2194, Japan.}
\address{$^2$\ Nakanihon Automotive College,
Gifu, 505-0077, Japan.}
\address{$^3$\ Institute for Theoretical Physics, University of Cologne, 
50923 K\"oln, Germany.}

\date{\today}
\begin{abstract} 
A new stochastic cellular automaton (CA) model of traffic flow, 
which includes slow-to-start effects and a driver's perspective,
is proposed by extending the Burgers CA and 
the Nagel-Schreckenberg CA model.
The flow-density relation of this model shows 
multiple metastable branches near the transition density
from free to congested traffic, which form a wide scattering area
in the fundamental diagram.
The stability of these branches and their velocity distributions
are explicitly studied by numerical simulations.
\end{abstract}
\section{Introduction}
Traffic problems have been attracting 
not only engineers but also physicists \cite{TGF99}.
Especially it has been widely accepted that 
the phase transition from free to congested
traffic flow can be understood using methods from
statistical physics \cite{CSS,He}.
In order to study the transition in detail, we need a realistic
model of traffic flow which should be minimal to clarify the underlying
mechanisms.
In recent years cellular automata (CA) \cite{Wo,CD} have been 
used extensively to study traffic flow in this context.
Due to their simplicity, CA models have also been applied
by engineers, e.g.\ for the simulation of complex traffic 
systems with junctions and traffic signals \cite{ACRI98}.

Many traffic CA models have been proposed so far \cite{CSS,FI,NS}, 
and among these CA, the deterministic rule-184 CA model (R184), 
which is one of the elementary CA classified by Wolfram \cite{Wo}, 
is the prototype of all traffic CA models.
R184 is known to represent the minimum movement of vehicles 
in one lane and shows a simple phase transition from free to 
congested state of traffic flow.
In a previous paper \cite{NT98}, 
using the ultra-discrete method \cite{TTMS}, the Burgers CA
(BCA) has been derived from the Burgers equation
\begin{equation}
 v_t = 2vv_x + v_{xx},
\label{CB}
\end{equation}
which was interpreted as a macroscopic traffic model \cite{MH}. 
The BCA is written using the minimum function $\min$ by
\begin{eqnarray}  \label{BCA}
  U_j^{t+1} = U_j^t &+& \min\{U_{j-1}^t, L - U_j^t\}
                   - \min\{U_j^t, L - U_{j+1}^t\},
\end{eqnarray}
where $U_j^t$ denotes the number of vehicles at the site $j$ 
and time $t$. 
If we put the restriction $L = 1$, it can be easily shown that the
BCA is equivalent to R184.
Thus we have clarified the connection between the Burgers equation 
and R184, which offers better understanding of the relation between 
macroscopic and microscopic models.

The BCA given above is considered as the {\it Euler} representation of 
traffic flow. As in hydrodynamics there is an another representation, 
called {\it Lagrange} representation \cite{Ni01}, which is specifically 
used for car-following models.
The Lagrange version of the BCA is given by \cite{MN}
\begin{equation}
 x_i^{t+1}=x_i^t+\min\{V_{max}, x^t_{i+S}-x^t_i-S\}, \label{lagBCA}
\end{equation}
where $V_{max}=S=L$ and $x^t_i$ is the position of $i$-th car at time $t$.
Note that in (\ref{lagBCA}) 
$S$ corresponds a ``perspective'' or anticipation parameter \cite{NT00} 
which represents the number of cars that a driver sees in front,
and $V_{max}$ is the maximum velocity of cars.
(\ref{lagBCA}) is derived from the BCA mathematically by using
an Euler-Lagrange (EL) 
transformation \cite{MN} which is a discrete version of the well-known
EL transformation in hydrodynamics.

In this paper, we will develop the BCA (\ref{lagBCA}) to a more realistic 
model by introducing slow-to-start (s2s) effects \cite{TT,BenJ,ASMS,BSSS} 
and a driver's perspective $S$.
Moreover, a stochastic generalization is also considered by combining it
with the Nagel-Schreckenberg (NS) model \cite{NS,SSNI}.

\section{Traffic models in Lagrange form}

First, let us extend (\ref{lagBCA}) to the case $V_{max}\ne S$ and 
combine it with the s2s model.
The s2s model \cite{Ni01} is written 
in Lagrange form as 
\begin{eqnarray}
  x_i^{t+1}&=&x_i^t+\min\{1, x^t_{i+1}-x^t_i-1,
x^{t-1}_{i+1}-x^{t-1}_i-1\}.
\label{eqS2S}
\end{eqnarray} 
Note that the inertia effect of cars is taken into account in this model.
Comparing (\ref{eqS2S}) and (\ref{lagBCA}), we see that, in the s2s
model, the velocity of a car depends not only on the present headway
$d_i^t=x^t_{i+1}-x^t_i-1$, but also on the past headway 
$d_i^{t-1}=x^{t-1}_{i+1}-x^{t-1}_i-1$. This rule has only a nontrivial
effect if $d_i^{t-1}=0$ and $d_i^{t}=1$, i.e.\ if the leading car has
started to move in the previous time step. In this case the following
car is not allowed to move immediately (s2s).

\begin{figure}[h] 
\begin{center} 
\includegraphics[width=0.8\textwidth]{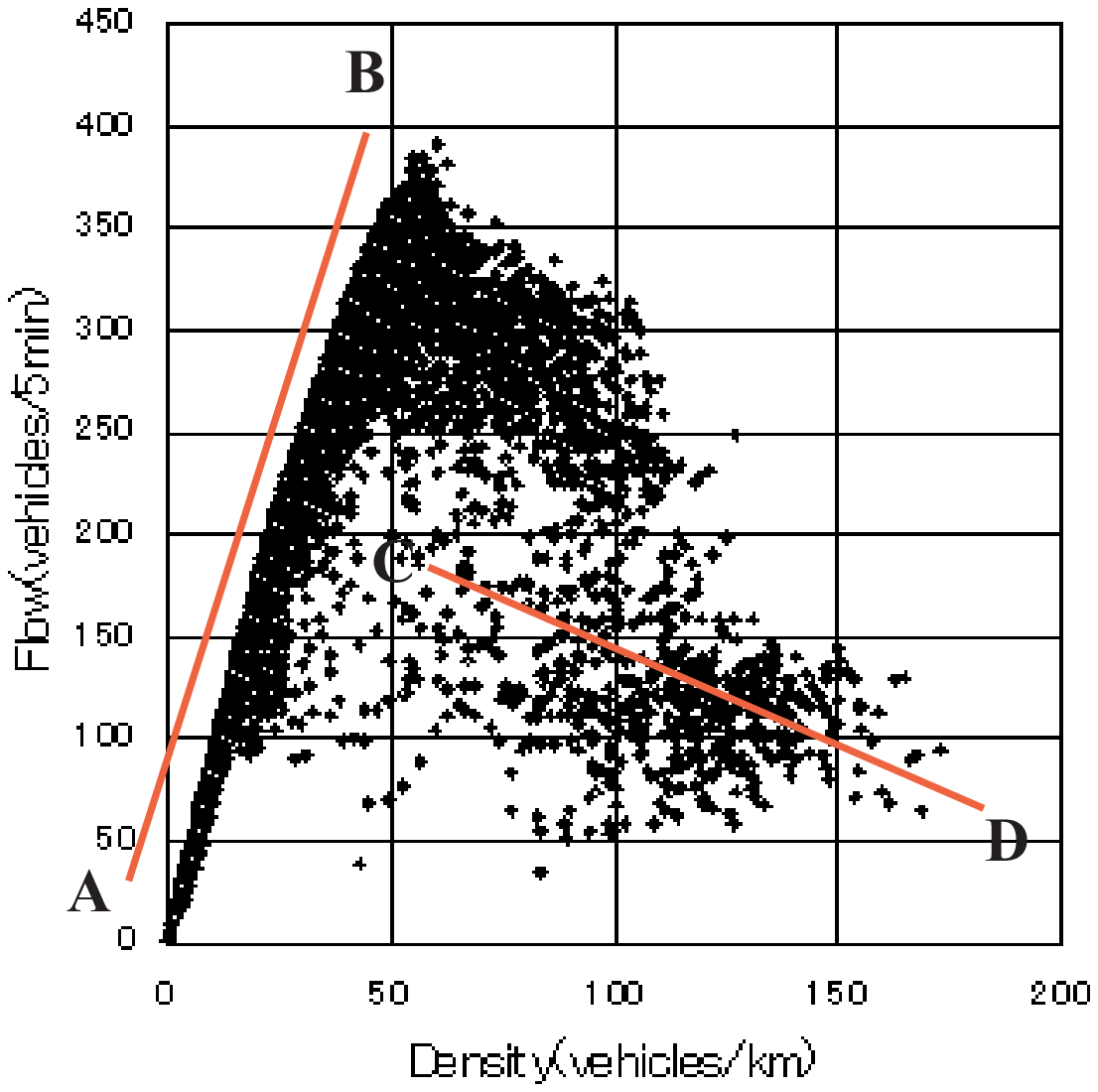}
\vspace{-1.5cm}
\end{center}
\caption{
An observed fundamental diagram at the Tomei expressway 
in Japan.
The gradient of the free line ($A-B$) and jamming line ($C-D$) 
is known to be about 100~km/hour and $-15$~km/hour.
We also see that there is a wide scattering area near the phase
transition region from free to jamming state. 
}
\label{fig1}
\end{figure}

Before combining (\ref{lagBCA}) and (\ref{eqS2S}), it is worth pointing 
out that we can choose the perspective parameter as 
$S=2$ in the model according to observed data.
We define the size of a cell as 7.5~m  and $V_{max}=5$ in our model 
according to the NS model. Since $V_{max}$ corresponds to about
100~km/hour in reality, then one time step in the CA model 
becomes 1.3~s.
Moreover, the gradient of the free line and jamming line in the fundamental
diagram, which is the dependence of the traffic flow $Q$ on
density $\rho$, is known to be about
100~km/hour and $-15$~km/hour \cite{He}
according to many observed data (see Fig.~\ref{fig1}) \cite{NH, THH}.
These values correspond to the typical free velocity and the
jam velocity, respectively.
Thus, considering the fact that 
the positive and negative gradient of each line 
is given by $V_{max}=5$ and 
$-S/2$, respectively, in the CA model \cite{Ni01}, we should choose
$S=1.5$ in the CA model. 
Since only integer numbers for $S$ are allowed in this model, 
we will simply choose $S=2$ for studying the effect of the perspective
of drivers. It is noted that other possibilities, such as 
velocity-dependent $S$ or stochastic choice of $S$, are also possible. 

Now by combining (\ref{lagBCA}) and (\ref{eqS2S}) 
we propose a new Lagrange model with $S=2$ 
which is defined by the rules listed below:
Let $v^{(0)}_i$ be the velocity of the $i$-th car at a time $t$. The update
procedure from $t$ to $t+1$ is divided into five stages:
\begin{enumerate}
 \item[1] {\it Accerelation}
\begin{equation}
 v^{(1)}_i = \min\{V_{max}, v^{(0)}_i+1\}. 
\label{acdet}
\end{equation} 
 \item[2] {\it Slow-to-accelerate effect}
\begin{equation}
 v^{(2)}_i = \min\{v^{(1)}_i, x^{t-1}_{i+2}-x^{t-1}_i-2\}. 
\label{s2a}
\end{equation}  
 \item[3] {\it Deceleration due to other vehicles}
\begin{equation}
 v^{(3)}_i = \min\{v^{(2)}_i, x^{t}_{i+2}-x^{t}_i-2\}.
\end{equation}  
 \item[4] {\it Avoidance of collision}
\begin{eqnarray}
 v^{(4)}_i &=& \min\{v^{(3)}_i, 
x^{t}_{i+1}-x^{t}_i-1 + v^{(3)}_{i+1}\}
\label{avoid}
\end{eqnarray}  
 \item[5] {\it Vehicle movement}
\begin{equation}
 x_i^{t+1} = x_i^t + v^{(4)}_i.
\end{equation}
\end{enumerate}
The velocity $v^{(4)}_i$ is used as $v^{(0)}_i$ in the next time step.
(\ref{avoid}) is the condition that the $i$-th car does not overtake
its preceding $(i+1)$-th car, including anticipation.
The accerelation (\ref{acdet}) is the same as in the NS model, which
is needed for a mild accerelating behaviour of cars.
In the step 2, we call (\ref{s2a}) as ``slow-to-accelerate'' 
instead of s2s. This is because this rule affects not only 
the behaviour of standing cars but also that of moving cars, which is
considered to be a generalization of s2s rule. 

\begin{figure}[h] 
\begin{center} 
\includegraphics[width=0.8\textwidth]{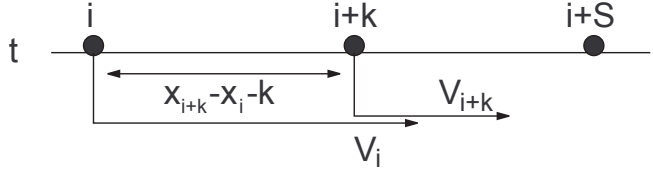}
\end{center}
\caption{
Collision-free condition between $i$-th and $i+k$-th car.
}
\label{fig2}
\end{figure}

It is not difficult to write down the new model in a single equation 
for general $S$. The result is
\begin{eqnarray}
 x_i^{t+1}&=&x_i^t+\min \left\{ V^t_i, 
 \min_{k=1,\cdots,S-1}(x^t_{i+k}-x^t_i-k
                                              +V^t_{i+k}) \right\},
\label{lagfm}
\end{eqnarray}
where the last term represents the collision-free condition explained
in Fig.~\ref{fig2}, and
\begin{eqnarray}
 V^t_i&=&\min\{V_{max}, x^{t-1}_{i+S}-x^{t-1}_i-S,
                    x^{t}_{i+S}-x^{t}_i-S, 
                     x^{t}_i-x^{t-1}_i+1\}.
\end{eqnarray}
The condition that there is no collision between the $i$-th and $i+k$-th 
cars ($k=1,\cdots,S-1$) is given by
\begin{equation}
 \hspace{-0.5cm} 
x^t_{i+k}-x^t_i-k + V^t_{i+k} \ge V^t_i,
\end{equation}
for $S \ge 2$ (if $S=1$ then we simply put $k=1$), which is identical
to the last term in (\ref{lagfm}).
In contrast to the NS model, the velocity of the preceeding
car is taken into account in the calculation of the safe velocity
in the step 4,
i.e.\ our model also includes anticipation effects.

\section{Metastable branches and their stability}
Next, we investigate the fundamental diagram of this new hybrid model.
\begin{figure}[h] 
\begin{center} 
\scalebox{0.7}{\includegraphics{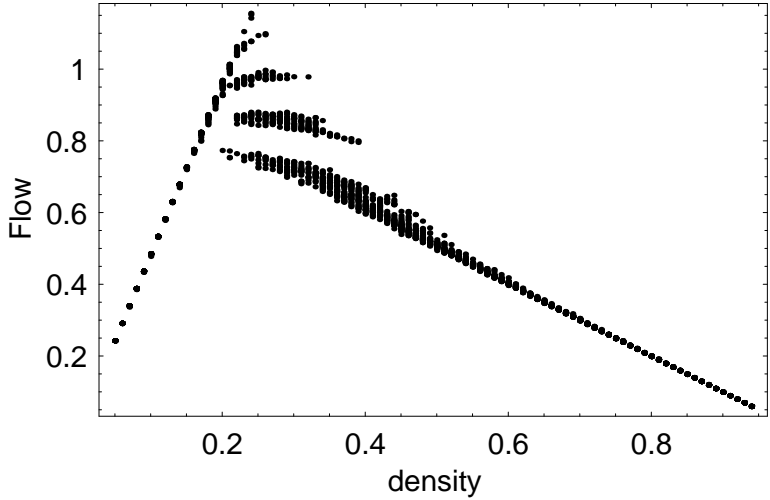}}
\end{center}
\caption{
Fundamental diagram of the new Lagrange model.
Parameters are set to
$V_{max}=5$ and $S=2$, and the spatial period is 100 sites.
The initial car density is varied from 0.05 to 0.95 in steps of 0.01.
At each density, we start calculations from 30 
randomly generated initial configurations, and show only the data
at the time $t=100$.
We observe several metastable branches in the deterministic case. 
The fluctuations of the branches show 
the fact that the asymptotic flow of the 
system sometimes becomes periodic instead of
stationary between 0.2 $\le\rho\le$ 0.5.
}
\label{fig3}
\end{figure}
In Fig.~\ref{fig3}, we observe a complex phase transition from a 
free to congested state near the critical density 0.2 $\sim$ 0.4.
There are many metastable branches in the diagram, similar to our
previous models in Euler form \cite{NT99,FNTI} 
or in other models with anticipation \cite{LdRS03}.
We also point out that there is a wide scattering area near the
critical density in the observed data (Fig.~\ref{fig1}) which may be 
related to these metastable branches. As we will discuss later, these
branches may account for some aspects of the scattering 
area observed empirically.

\begin{figure}[h] 
\begin{center} 
\includegraphics[width=0.4\textwidth]{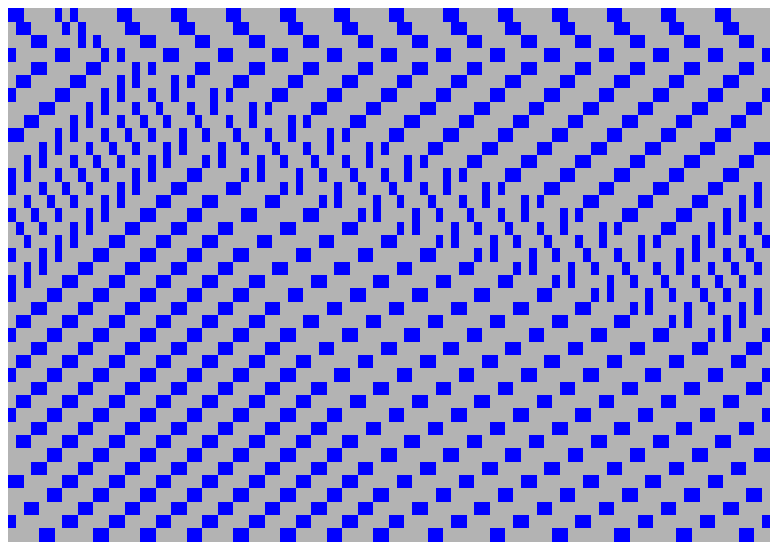}
\includegraphics[width=0.4\textwidth]{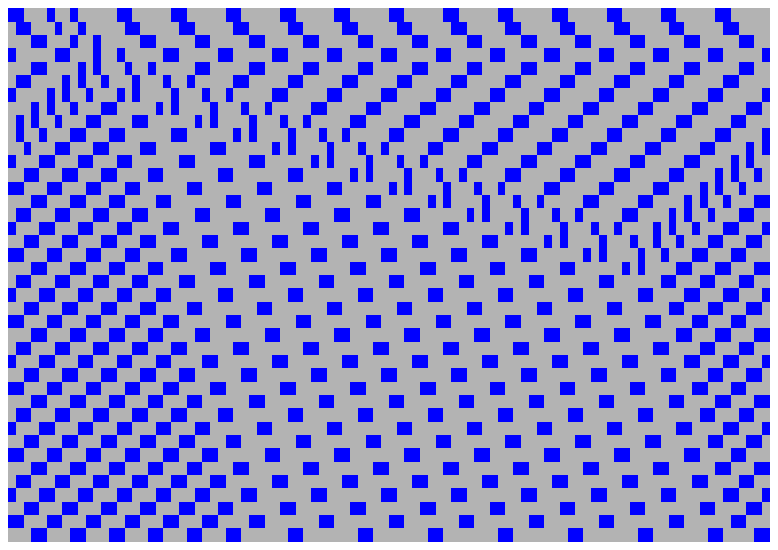}\\
(a)\hspace{4cm}(b)\\
\includegraphics[width=0.4\textwidth]{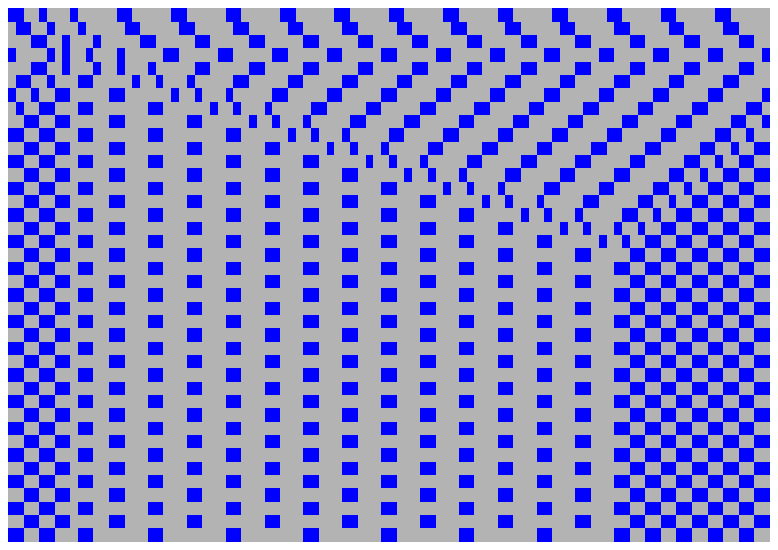}
\includegraphics[width=0.4\textwidth]{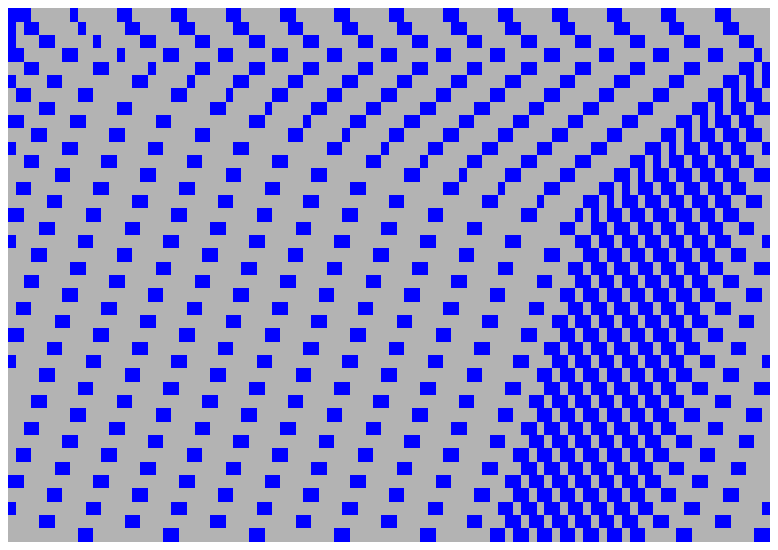}\\
(c)\hspace{4cm}(d)\\
\includegraphics[width=0.4\textwidth]{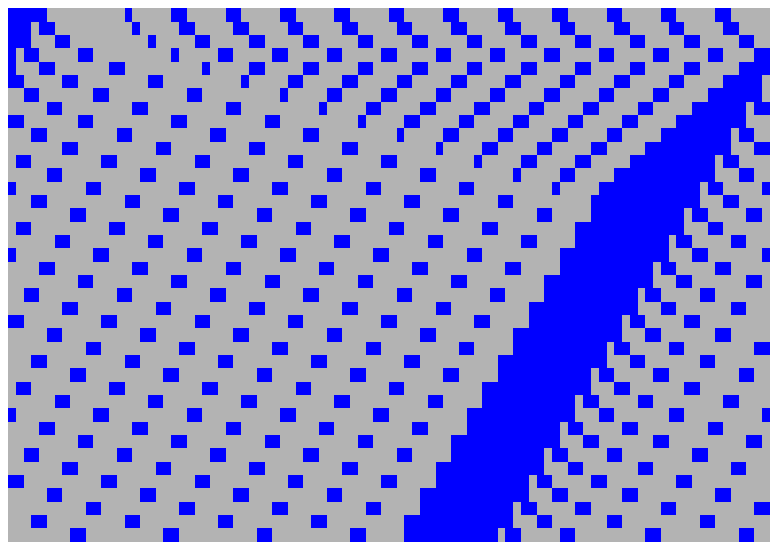}\\
(e)
\end{center}
\caption{
Spatio-temporal patterns of evolution of the uniform flow 
$\cdots 11000001100000 \cdots$ at density $\rho=2/7$
with different strength of perturbation:
(a) very weak, (b) less weak, (c) medium, (d) stronger and finally (e) 
strongest perturbation. The details of these perturbations are 
all explained in detail in the text.
The stationary state of these five cases correspond to a state in 
each metastable branch appearing in Fig.~\ref{fig3}, although the branch
corresponding to (a) does not appear in the numerical simulations
with random initial conditions.
}
\label{fig4}
\end{figure}

\begin{figure}[h] 
\begin{center} 
\includegraphics[width=0.4\textwidth]{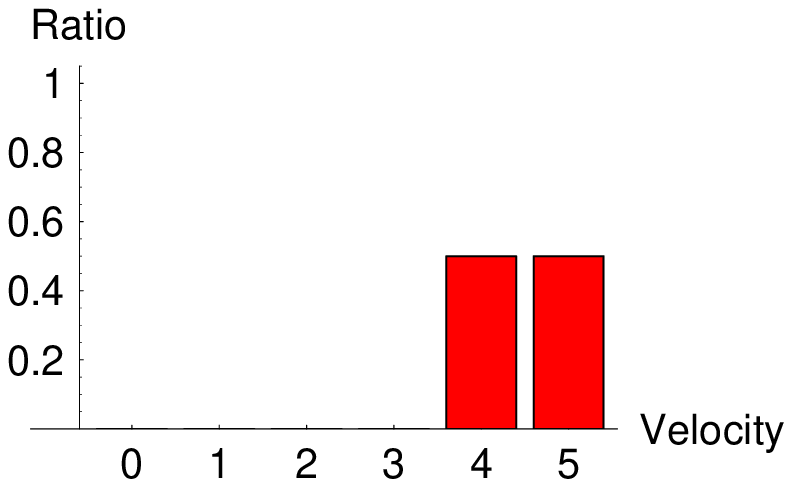}
\includegraphics[width=0.4\textwidth]{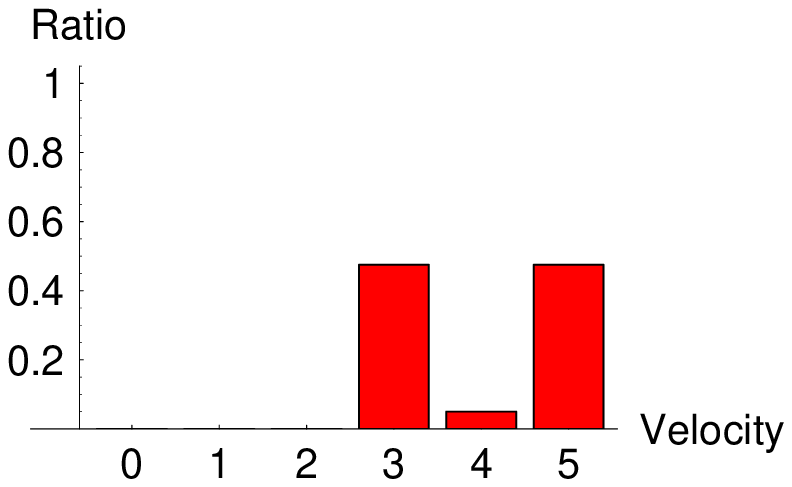}\\
(a)\hspace{4cm}(b)\\
\includegraphics[width=0.4\textwidth]{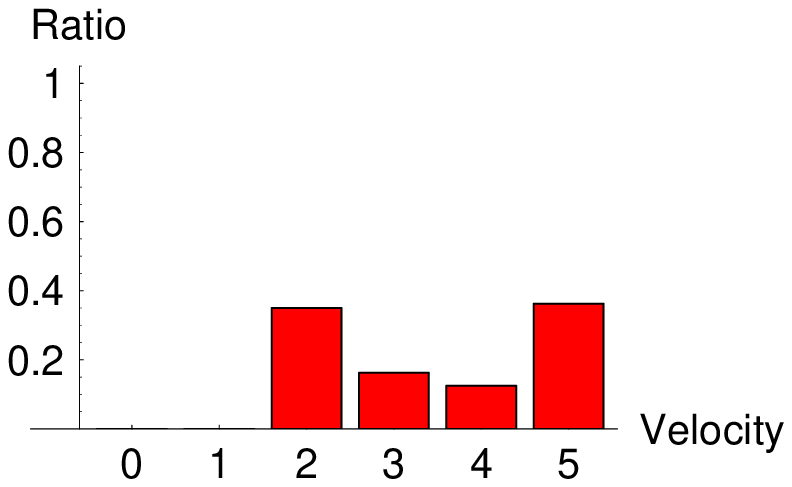}
\includegraphics[width=0.4\textwidth]{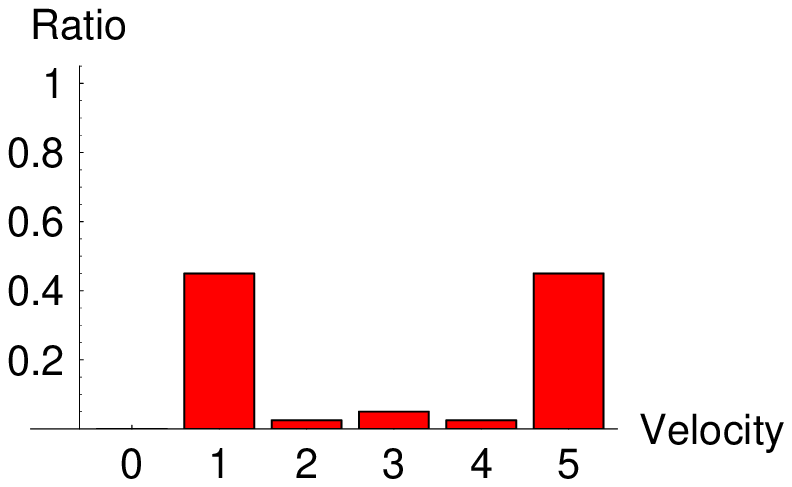}\\
(c)\hspace{4cm}(d)\\
\includegraphics[width=0.4\textwidth]{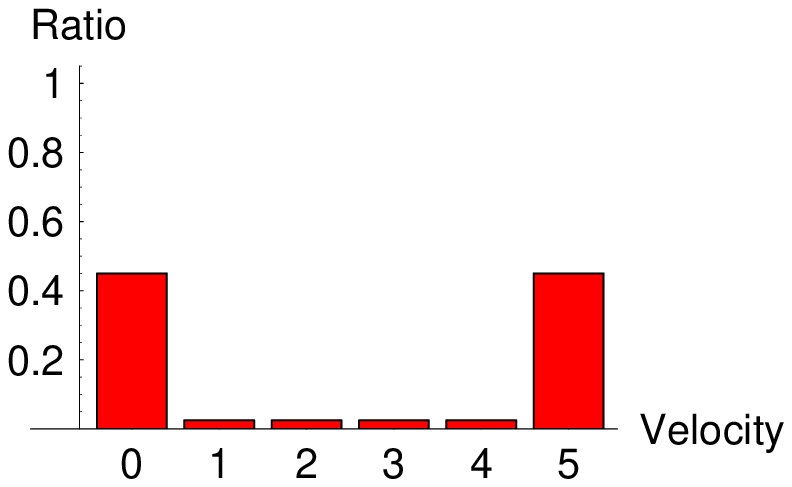}\\
(e)
\end{center}
\caption{
Velocity distributions in the case $\rho=2/7$, corresponding 
to the states given in fig.~\ref{fig4}.
}
\label{fig5}
\end{figure}

First, we discuss properties of the state in the metastable
branches.
In all cases it consists of {\em pairs of vehicles} that move coherently
with vanishing headway (see Fig.~\ref{fig4}). 
Cars are represented by black squares, and the direction of the road is
horizontal right and time axis is vertical down. 
The corresponding velocity distributions are also given in Fig.~\ref{fig5}.
We see that there are stopping cars which velocity are zero only
in the case of the lowest branch given in the state in Fig.~\ref{fig4} (e).

\begin{figure}[h] 
\begin{center} 
\includegraphics[width=0.6\textwidth]{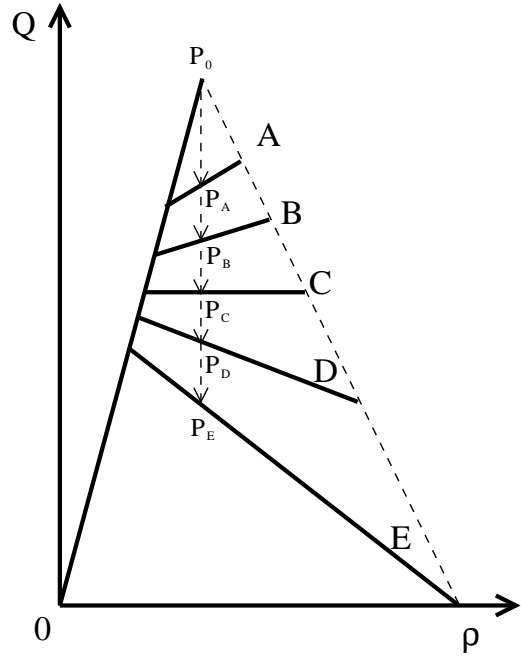}
\end{center}
\caption{
A schematic diagram of the metastable branches ($A,B,C$ and $D$) 
and the jamming line ($E$) in the new model. 
The highest flow state is represented by $P_0$, which is quite
unstable and easy to go down to the lower flow state
$P_A, \cdots, P_E$ according to the magnitude of the perturbation.
}
\label{fig6}
\end{figure}

Next let us calculate the flow-density relation for each branch. 
In the metastable branches
we find phase separation into a free-flow and a jamming region.
In the former, pairs move with velocity $v_f$ and a headway of $d_f$
empty cells between consecutive pairs. In the jammed region, the velocity
of the pairs is $v_j$ and the headway $d_j$.
$N_j$ and $N_f$ are the numbers of cars in the jamming cluster 
and the free uniform flow, respectively. We assume $N_f$ and $N_j$
to be even so that there are $N_f/2$ and $N_j/2$ pairs, respectively.
Then the total number of cars $N$ is given by $N=N_j+N_f$ and the total length
of the system becomes $l=(d_j+2)N_j/2+(d_f+2)N_f/2$.
Since the average velocity is $\bar{v}=(N_fv_f+N_jv_j)/N$ and
density and flow of the system 
are given by $\rho=N/l$ and $Q=\rho \bar{v}$,
we obtain the flow-density relation as
\begin{equation}
 Q=2\frac{v_f-v_j}{d_f-d_j}+\left( v_j - (d_j+2)\frac{v_f-v_j}{d_f-d_j}
   \right)\rho.\quad
\end{equation}
From the stationary states in Fig.~\ref{fig4} we have
\begin{eqnarray}
&&(a): (v_f,v_j,d_f,d_j)=(5,4,6,4)\nonumber\\
&&(b): (v_f,v_j,d_f,d_j)=(5,3,7,3)\nonumber\\
&&(c): (v_f,v_j,d_f,d_j)=(5,2,8,2)\nonumber\\
&&(d): (v_f,v_j,d_f,d_j)=(5,1,9,1)\nonumber\\
&&(e): (v_f,v_j,d_f,d_j)=(5,0,10,0)
\end{eqnarray}
Therefore the resulting equations for each branch are
\begin{equation}
 Q=1+c\rho
\label{metaQ}
\end{equation}
where $c=1,1/2,0,-1/2,-1$, which correspond to
the branches $A,B,C,D$ and $E$ in Fig.~\ref{fig6}, respectively.
End points of the branches are found to be given by
$\rho_1=2/(d_f+2)$ and $\rho_2=2/(d_j+2)$, where 
$\rho_1$ is the point that the metastable 
branches intersect the free flow branch,
and $\rho_2$ is the maximal possible density in the metastable branches.
Note that all $\rho_2$ lie on the line $Q=-2\rho+2$, which is indicated
as the broken line in Fig.~\ref{fig6}.

Next let us now study the stability of each metastable branch.
We mainly consider the density $\rho=2/7$ and,
in particular, we will focus on the uniform flow represented by
$\cdots 11000001100000 \cdots$, which shows the highest flow given in 
the point $P_0$ in Fig.~\ref{fig6}.

Spatio-temporal patterns due to various kinds of perturbations
are already seen in Fig.~\ref{fig4}. Perturbation in this case
means that some cars are shifted backwards at the initial configuration.
The initial conditions for Fig.~\ref{fig4}(a)-(e) are given as follows:
\begin{enumerate}
 \item[(a)] Very  weak perturbation (one car is shifted one site
            backwards)$\cdots 11000010100000 \cdots$
 \item[(b)] Weak perturbation (one car is shifted two sites
            backwards)$\cdots 11000100100000 \cdots$
 \item[(c)] Moderate perturbation (one car is shifted three sites
            backwards)$\cdots 11001000100000 \cdots$
 \item[(d)] Strong perturbation (one car is shifted five sites
            backwards)$\cdots 11100000100000 \cdots$
 \item[(e)] Strongest perturbation (three cars are shifted backwards)
$\cdots 11111000000000 \cdots$
\end{enumerate}
The stationary state of (a), (b), (c), (d) 
and (e) are given by the points
$P_A$, $P_B$, $P_C$, $P_D$ and $P_E$ in Fig.~\ref{fig6}, respectively.
That is, if the system in $P_0$ is perturbated, then the flow
easily goes down to a lower branch in the course of time depending on
the magnitude of the perturbation.
Since the density does not change due to the perturbation,
we obtain $P_A:(2/7,9/7)$, $P_B:(2/7,8/7)$,
$P_C:(2/7,1)$, $P_D:(2/7,6/7)$ and $P_E:(2/7,5/7)$ by substituting
$\rho=2/7$ into eq.~(\ref{metaQ}).

We see a jamming cluster propagating backwards
in the cases of (d) and (e) in Fig.~\ref{fig4}.
In other cases the jamming cluster propagates forward ((a) and (b)) 
or does not move (c).
These facts are related to the gradient of the metastable 
branches which are given by $c$ according to eq.~(\ref{metaQ}).

\section{Stochastic generalization}
Finally we will combine the above model with the NS model
in order to take into account the randomness of drivers.
The NS model is written in Lagrange form as
\begin{eqnarray}
  x_i^{t+1}=x_i^t+\max\left\{0,\min\{V, x^t_{i+1}-x^t_i-1,
x^t_i-x^{t-1}_i+1\}-\eta^t_i\right\}.
\label{eqns}
\end{eqnarray}
where $\eta^t_i=1$ with probability $p$ and  $\eta^t_i=0$ with 
probability $1-p$. The last term in the mininum in (\ref{eqns}) 
represents the acceleration of cars.
The randomness in this model is considered as a kind of random braking
effect, which is known to be responsible for spontaneous jam
formation often observed in real traffic \cite{CSS}. We also 
consider random {\it accerelation} in this model which is not taken into 
account in the NS model.

Thus a stochastic generalization of the hybrid model in the case of $S=2$ 
is similarly given by the following set of rules:
\begin{enumerate}
 \item[1] {\it Random accerelation}
\begin{equation}
 v^{(1)}_i = \min\{V_{max}, v^{(0)}_i+\eta_a\}. 
\end{equation} 
where $\eta_a=1$ with the probability $p_a$ and 
$\eta_a=0$ with $1-p_a$.\\
 \item[2] {\it Slow-to-accelerate effect}
\begin{equation}
 v^{(2)}_i = \min\{v^{(1)}_i, x^{t-1}_{i+S}-x^{t-1}_i-S\}.
\end{equation}  
 \item[3] {\it Deceleration due to other vehicles}
\begin{equation}
 v^{(3)}_i = \min\{v^{(2)}_i, x^{t}_{i+S}-x^{t}_i-S\}.
\end{equation} 
 \item[4] {\it Random braking}
\begin{equation}
 v^{(4)}_i = \max\{v^{(3)}_i-\eta_b, 0\}
\end{equation}  
where $\eta_b=1$ with the probability $p_b$ and 
$\eta_b=0$ with $1-p_b$.\\
 \item[5] {\it Avoidance of collision}
\begin{eqnarray}
 v^{(n+1)}_i &=& \min\{v^{(n)}_i, 
x^{t}_{i+1}-x^{t}_i-1 + v^{(n)}_{i+1}\}
\end{eqnarray}  
with $n \ge 4$, which is an iterative equation that has to be applied 
until $v$ converges to $v^{(n+1)}_i = v^{(n)}_i (\equiv v_i)$.\\
 \item[6] {\it Vehicle movement}
\begin{equation}
 x_i^{t+1} = x_i^t + v_i.
\end{equation}
\end{enumerate}
Again the velocity $v_i$ is used as $v^{(0)}_i$ in the next time step.
Step 5 must be applied to each car iteratively until its velocity
does not change any more, which ensures that this model is free from
collisions. This is the difference between the deterministic and
stochastic case. In the deterministic model it is sufficient to apply 
the avoidance of collision stage only once in each update, while in 
the stochastic case generically it has to be applied a few times in order 
to avoid collisions between successive cars. 

\begin{figure}[h] 
\begin{center} 
\includegraphics[width=0.4\textwidth]{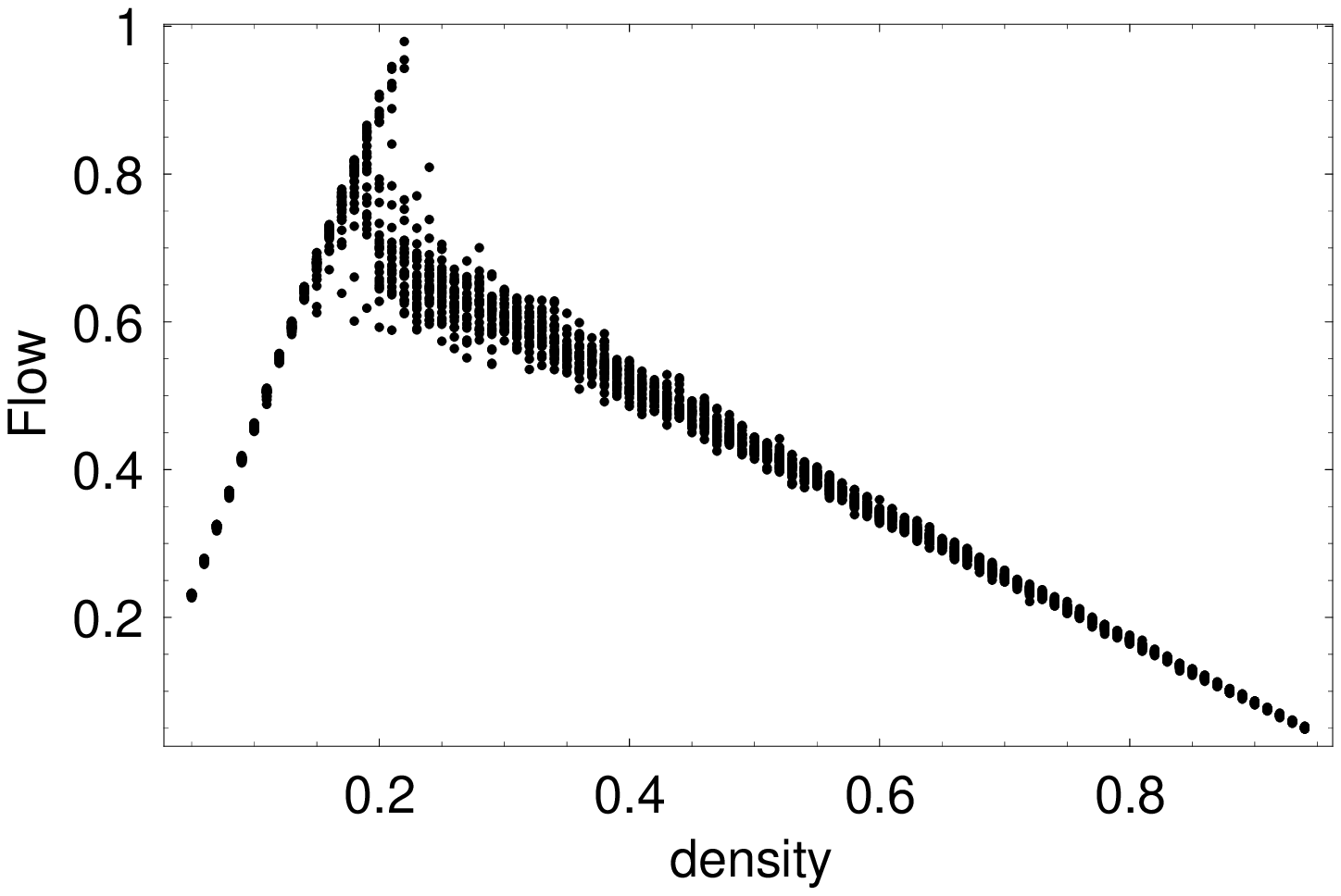}
\includegraphics[width=0.4\textwidth]{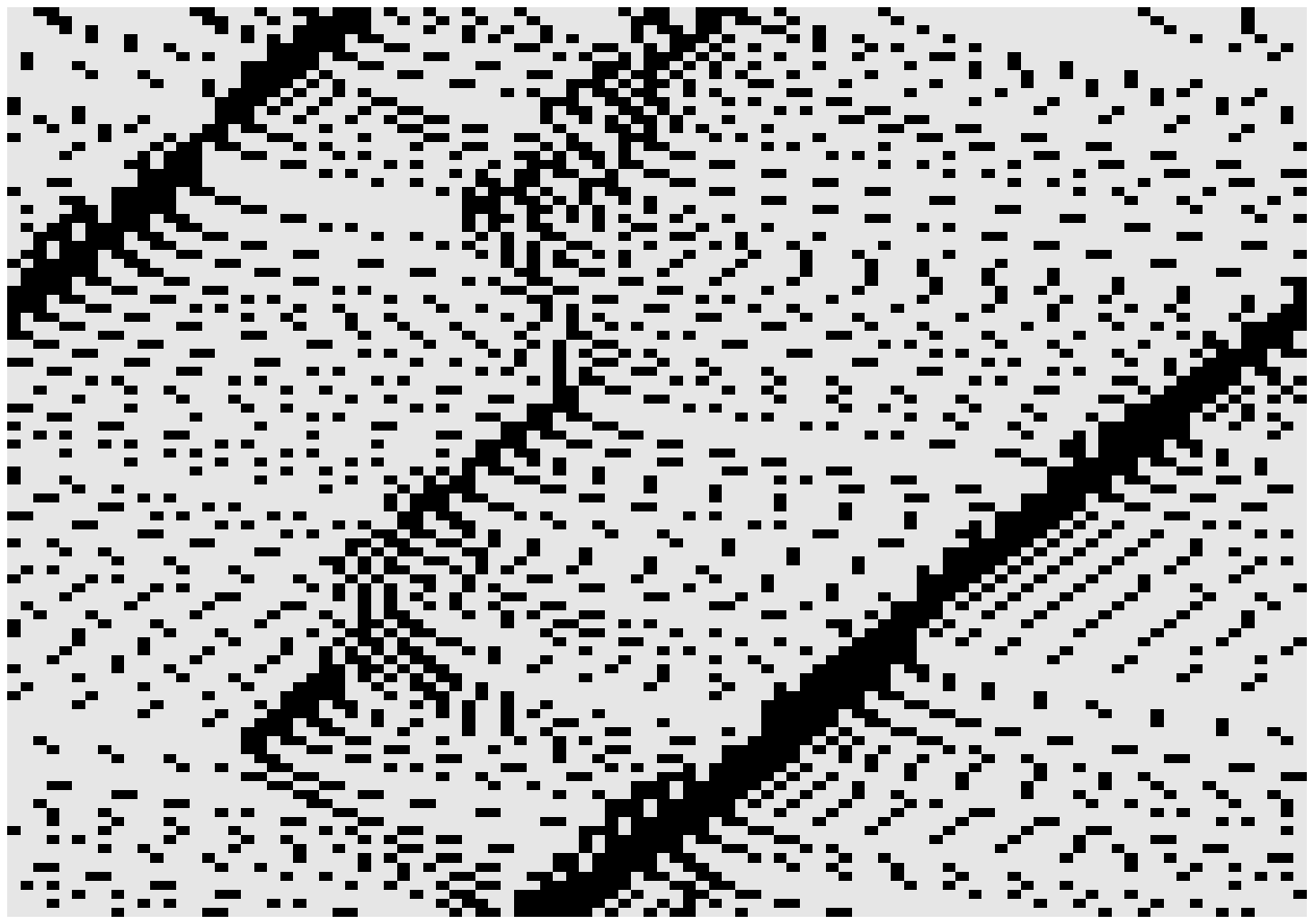}\\
\includegraphics[width=0.4\textwidth]{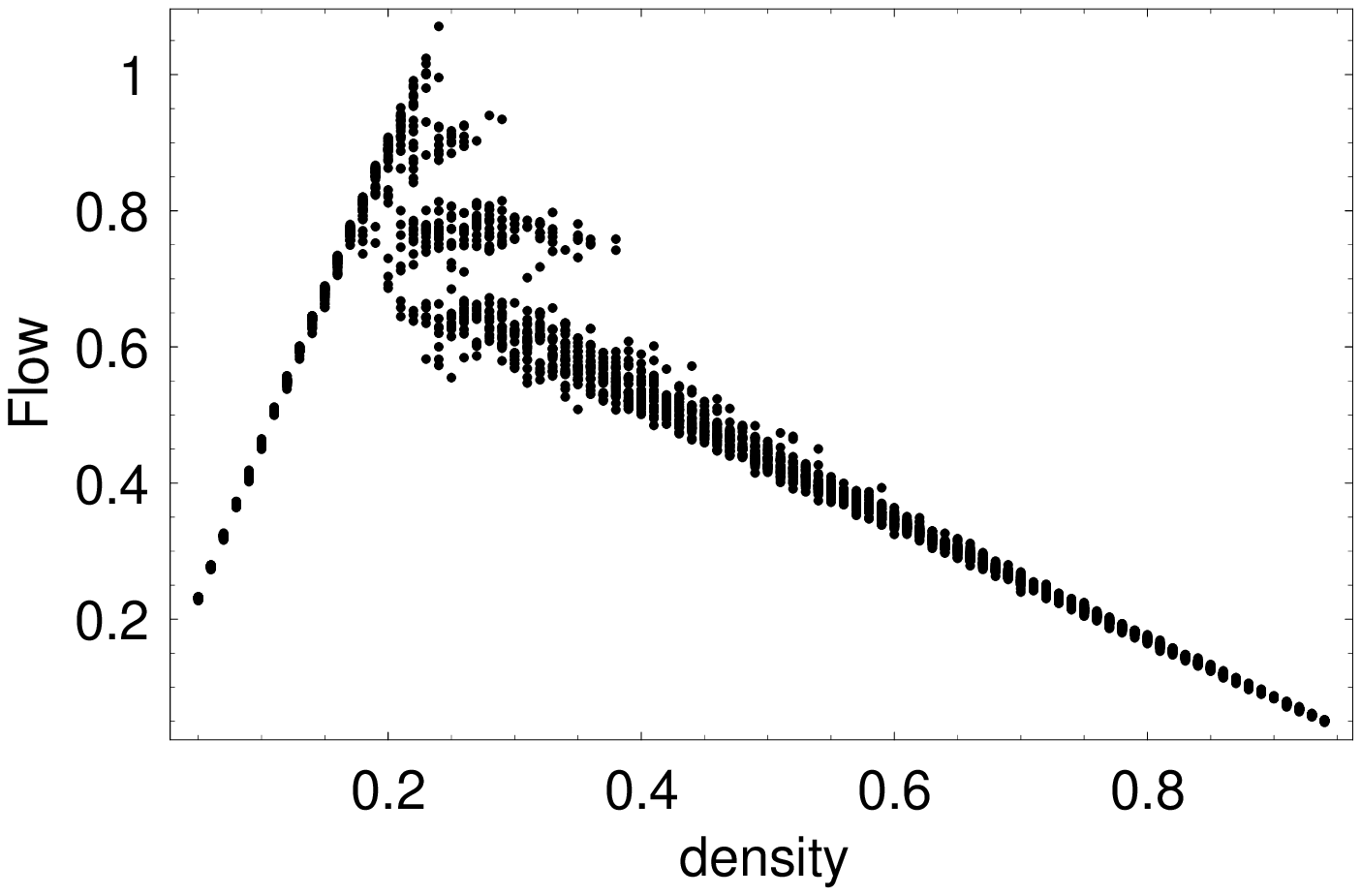}
\includegraphics[width=0.4\textwidth]{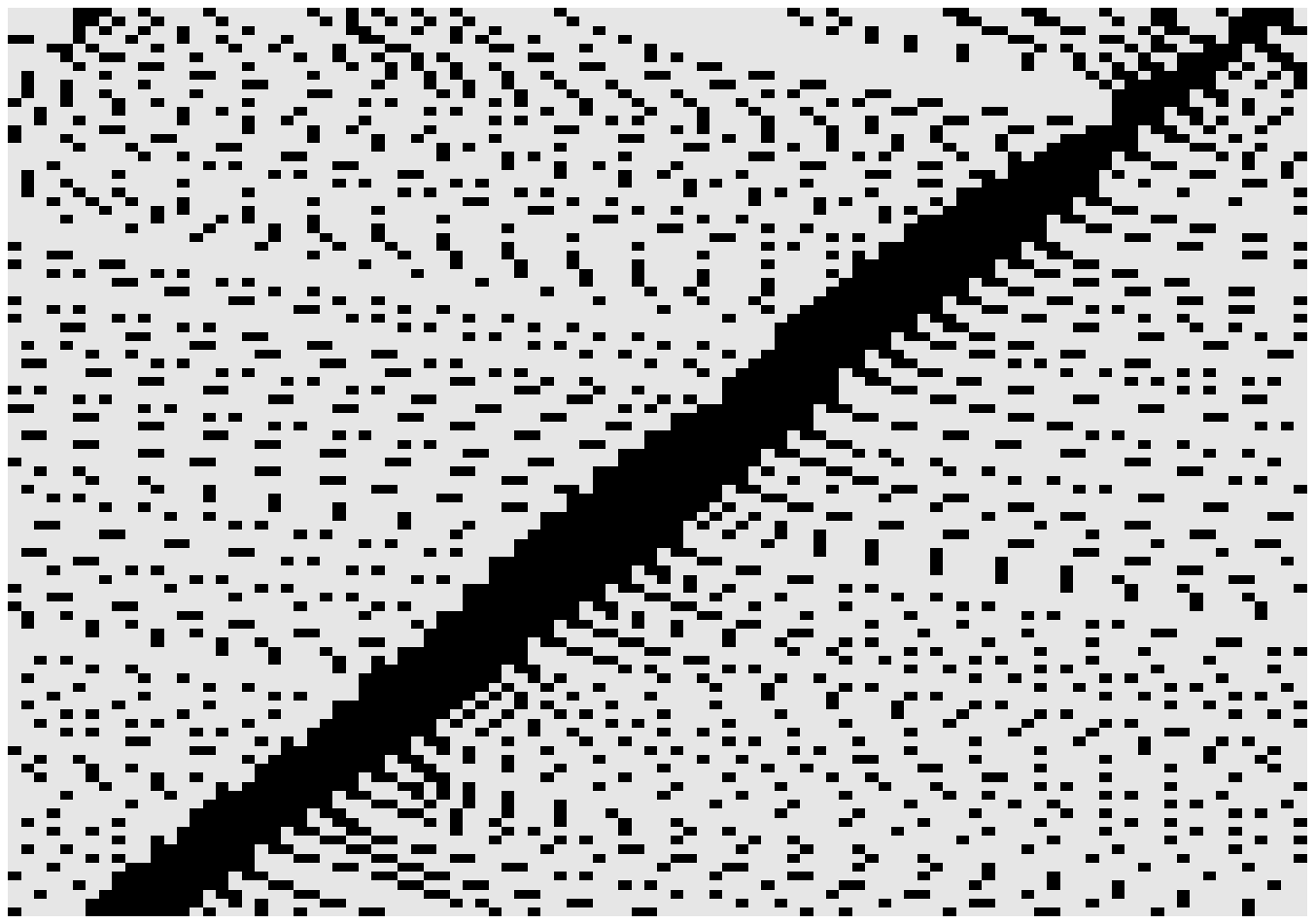}\\
\includegraphics[width=0.4\textwidth]{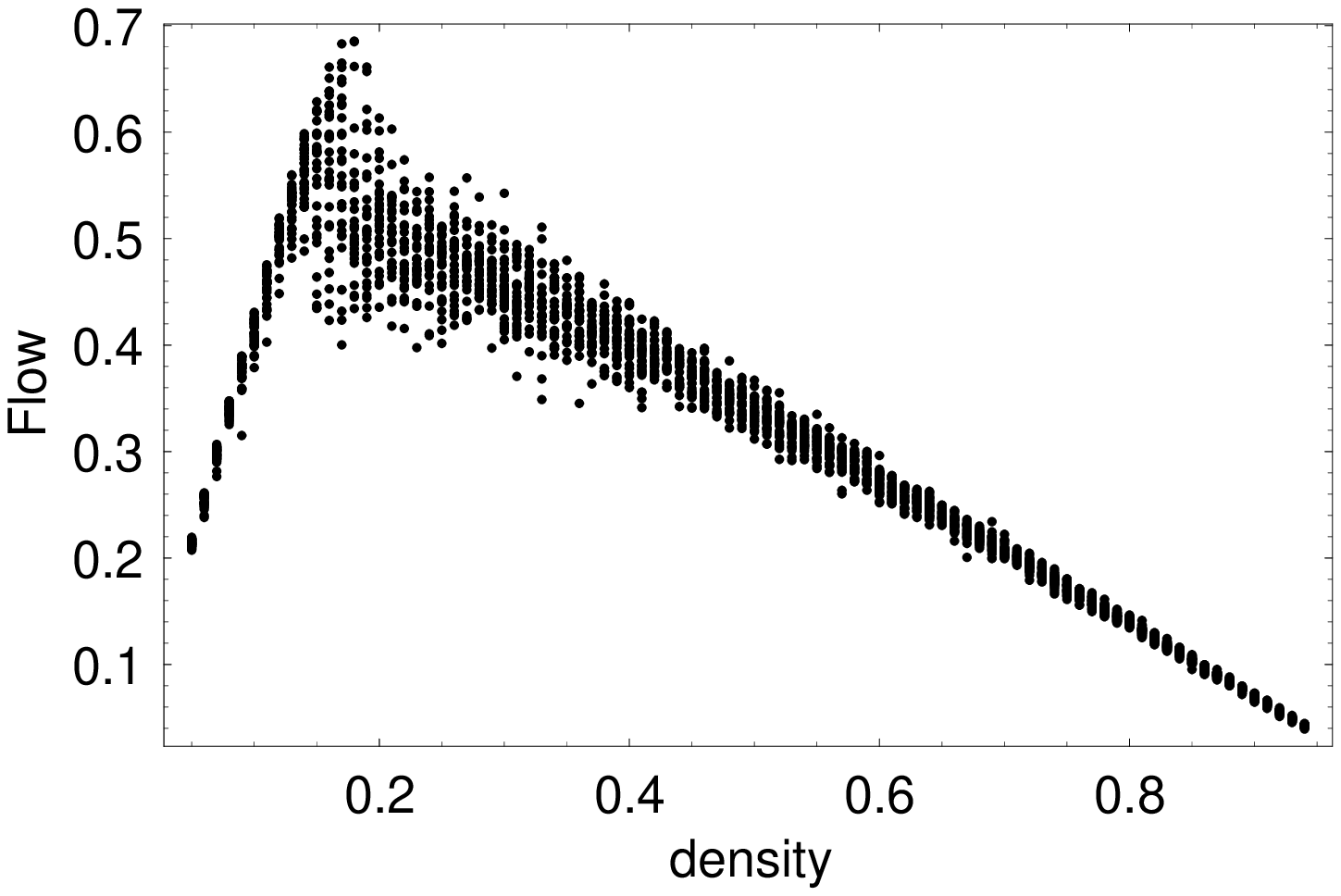}
\includegraphics[width=0.4\textwidth]{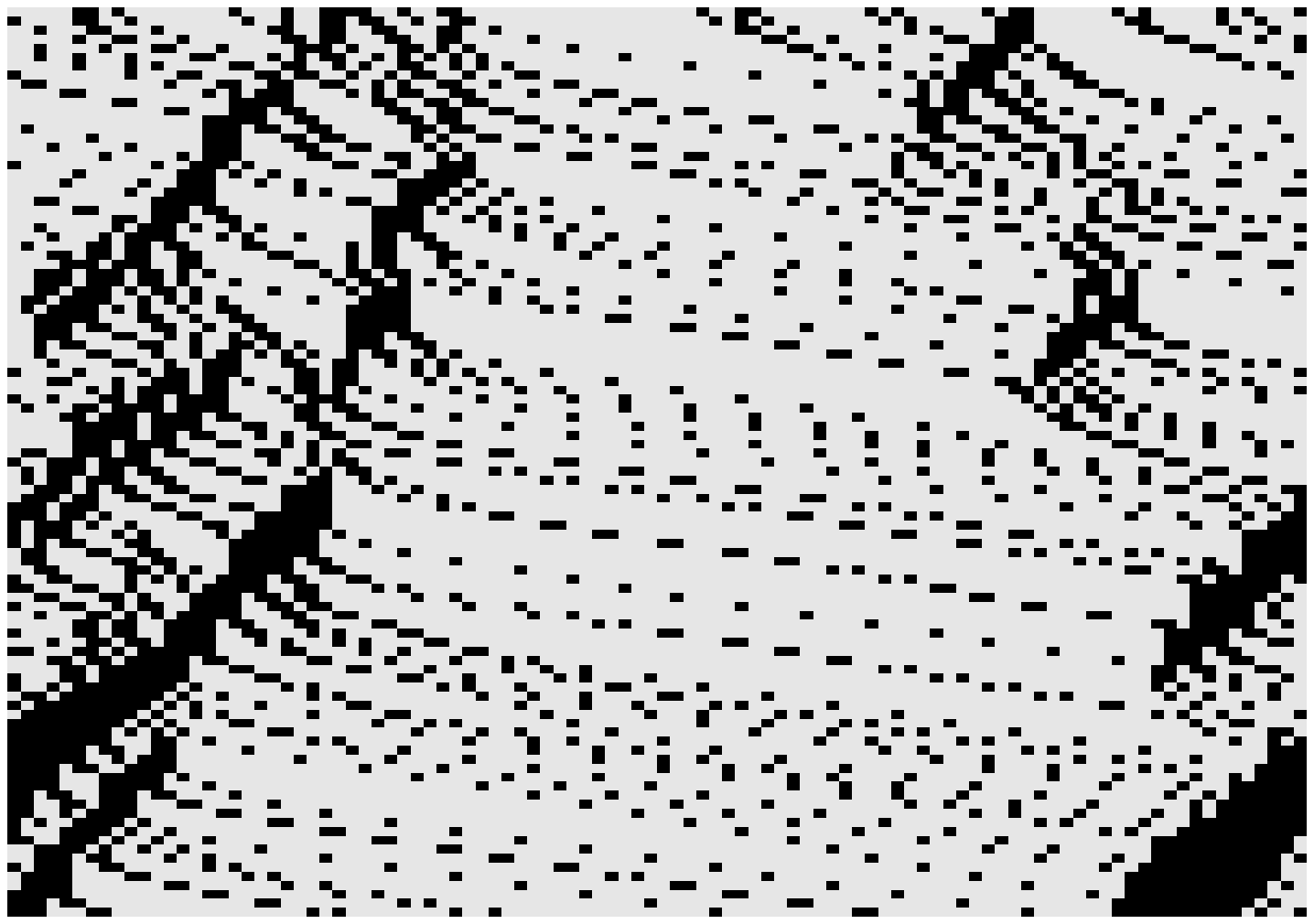}
\end{center}
\caption{
Fundamental diagrams and typical spatio-temporal patterns 
of the new stochastic model with different value of 
random parameters. 
Parameters are set to $V_{max}=5$ and $S=2$.  
Upper two figures are the case of $p_a=0$ and $p_b=0.2$,
middle ones are $p_a=0.8$ and $p_b=0$, and the bottom ones 
are $p_a=0.8$ and $p_b=0.2$.
}
\label{fig7}
\end{figure}
The fundamental diagrams of this stochastic model for
some values of $p_a$ and $p_b$ are given in Fig.~\ref{fig7}.
The randomization effect can be considered 
as a sort of perturbation to the deterministic model. 
Hence some unstable branches seen in the deterministic case 
disappear in the stochastic case, especially if we consider the
random braking effect as seen in Fig.~\ref{fig7}.
Random accerelation itself 
does not significantly destroy the metastable branches.
Moreover, from the spatio-temporal pattern
it is found that 
spontaneous jam formation is observed only if we allow
random braking. Random accerelation alone 
is not sufficient to produce spontaneous jamming.
We also note that a wider scattering area appears if we introduce
both random accerelation and braking. 
\section{Concluding discussions}
In this paper we have proposed a new hybrid model of traffic flow
of Lagrange type
which is a combination of the BCA and the s2s model. Its stochastic
extension is also proposed by further incorporating stochastic
elements of the NS model and
random accerelation. The model shows several metastable branches around 
the critical density in its
fundamental diagram. The upper branches are
unstable and will decrease its flow under perturbations. It is shown 
that the magnitude of a perturbation determines the final value of flow 
in the stationary state. 
Moreover, introduction of stochasticity in the model
makes the metastable branches dilute and hence produces a wide scattering
area in the fundamental diagram.
We would like to point out that this metastable 
region around the phase transition density is similar to so-called 
synchronized flow proposed by \cite{KR}.
Our investigation shows that one possible origin of such a
region is the occurance of many intermediate congested states near 
the critical density.
If some of them are unstable due to perturbation or randomness,
then a dense scattering area near the critical density is formed 
around the metastable branches.
This is in some sense in between the two cases of a 
fundamental diagram based approach (with unique flow-density relation)
and the so-called 3-phase model of \cite{KKW02} which exhibits a 
full two-dimensional region of allowed states even in the deterministic
limit.

\section*{Acknowledgment} 
This work is supported in part by a Grant-in-Aid from the Japan Ministry of
 Education, Science and Culture.



\section*{References}
\begin{thebibliography}{99}

 \bibitem{TGF99}
D. Helbing and H. J. Herrmann and M. Schreckenberg and 
D. E. Wolf (eds.),
"Traffic and Granular Flow '99",
(Springer, 2000, Berlin).

 \bibitem{CSS}
D. Chowdhury, L. Santen and A. Schadschneider, 
Phys.\ Rep. {\bf 329} (2000) 199.

 \bibitem{He}
D. Helbing, Rev.\ Mod.\ Phys., {\bf 73} (2001) 1067.

 \bibitem{Wo}
S. Wolfram,
{\em Theory and applications of cellular automata},
(World Scientific, 1986, Singapore).

 \bibitem{CD}
B. Chopard and M. Droz,
{\em Cellular Automata Modeling of Physical Systems},
(Cambridge University Press, 1998).

\bibitem{ACRI98}
  S. Bandini, R. Serra and F. S. Liverani (eds.),
  {\em Cellular Automata: Research Towards Industry},
  (Springer, 1998).

 \bibitem{FI}
M. Fukui and Y. Ishibashi, J.\ Phys.\ Soc.\ Jpn. {\bf 65} (1996) 1868.

 \bibitem{NS}
K. Nagel and M. Schreckenberg, J.\ Phys.\ I France {\bf 2} (1992) 2221.

 \bibitem{NT98}
K. Nishinari and D. Takahashi, J.\ Phys.\ A. {\bf 31} (1998) 5439.

 \bibitem{TTMS}
T. Tokihiro, D. Takahashi, J. Matsukidaira, and J. Satsuma,
Phys.\ Rev.\ Lett. {\bf 76} (1996) 3247.

 \bibitem{MH}
T. Musya and H. Higuchi, J.\ Phys.\ Soc.\ Jpn. {\bf 17} (1978) 811.

 \bibitem{Ni01}
K. Nishinari, J.\ Phys.\ A {\bf 34} (2001) 10727.

 \bibitem{MN}
J. Matsukidaira and K. Nishinari, Phys.\ Rev.\ Lett. {\bf 90} (2003) 088701.

 \bibitem{NT00}
K. Nishinari and D. Takahashi, J.\ Phys.\ A {\bf 33} (2000) 7709.

 \bibitem{TT}
M. Takayasu and H. Takayasu, Fractals {\bf 1} (1993) 860.

\bibitem{BenJ}
S.C.\ Benjamin and N.F.\ Johnson, J.\ Phys.\ A {\bf 29} (1996) 3119.

\bibitem{ASMS}
A.~Schadschneider and M.~Schreckenberg, Ann.\ Physik {\bf 6} (1997) 541.

\bibitem{BSSS}
R.\ Barlovic, L.\ Santen, A.\ Schadschneider, and M.\ Schreckenberg,
Eur.\ Phys.\ J.\ {\bf 5} (1998) 793.

\bibitem{SSNI}
M.~Schreckenberg, A.~Schadschneider, K.~Nagel, and N.~Ito, Phys.\ Rev.\ E
{\bf 51} (1995) 2939.

 \bibitem{NH}
K. Nishinari and M. Hayashi,
{\em Traffic statistics in Tomei express way},
(The Mathematical Society of Traffic Flow, 1999, Nagoya).

 \bibitem{THH}
M. Treiber, A. Hennecke and D. Helbing, Phys.\ Rev.\ E, {\bf 62}
   (2000) 1805.

 \bibitem{NT99}
K. Nishinari and D. Takahashi,
  J.\ Phys.\ A., {\bf 32} (1999) 93.

 \bibitem{FNTI}
M. Fukui, K. Nishinari and D. Takahashi and Y. Ishibashi,
   Physica\ A, {\bf 303} (2002) 226.

 \bibitem{LdRS03}
M.E. Larraga, J.A. del Rio and A. Schadschneider,
   (2003) cond-mat/0306531.

 \bibitem{KR}
B. S. Kerner and H. Rehborn, Phys.\ Rev.\ E, {\bf 53} (1996)  1297.

\bibitem{KKW02}
B.S.~Kerner, S.L.\ Klenov, and D.E.\ Wolf, J.\ Phys.\ A {\bf 35} (2002) 9971. 

\end{thebibliography}
\end{document}